\documentclass[sigconf]{acmart}

\AtBeginDocument{%
  }

\usepackage{microtype}
\usepackage{graphicx}

\usepackage{booktabs}
\usepackage{multirow}
\usepackage{enumitem}

\usepackage{CJK}
\usepackage{color}
\usepackage{verbatim}
\usepackage{threeparttable}

\setcopyright{acmcopyright}
\copyrightyear{2018}
\acmYear{2018}
\acmDOI{XXXXXXX.XXXXXXX}

\acmConference[Conference acronym 'XX]{Make sure to enter the correct
  conference title from your rights confirmation emai}{June 03--05,
  2018}{Woodstock, NY}
\acmPrice{15.00}
\acmISBN{978-1-4503-XXXX-X/18/06}

\begin{document}

\title{Modeling Multi-interest News Sequence for News Recommendation}

\author{Rongyao Wang}
\affiliation{%
	\institution{School of Computer, Qilu University of Technology (Shandong Academy of Sciences)}
	\streetaddress{1 Th{\o}rv{\"a}ld Circle}
	\city{Jinan}
	\country{China}}
\email{honorw@foxmail.com}

\author{Wenpeng Lu}
\authornote{Corresponding author.}
\affiliation{%
	\institution{School of Computer, Qilu University of Technology (Shandong Academy of Sciences)}
	\city{Jinan}
	\country{China}}
\email{lwp@qlu.edu.cn}

\renewcommand{\shortauthors}{Wang et al.}

\begin{abstract}
A session-based news recommender system recommends the next news to a user by modeling the potential interests embedded in a sequence of news read/clicked by her/him in a session. Generally, a user's interests are diverse, namely there are multiple interests corresponding to different types of news, e.g., news of distinct topics, within a session. 
However, most of existing methods typically overlook such important characteristic and thus fail to distinguish and model the potential multiple interests of a user, impeding accurate recommendation of the next piece of news. Therefore, this paper proposes \textit{multi-interest news sequence} (MINS) model for news recommendation. In MINS, a news encoder based on self-attention is devised on learn an informative embedding for each piece of news, and then a novel parallel interest network is devised to extract the potential multiple interests embedded in the news sequence in preparation for the subsequent next-news recommendations. The experimental results on a real-world dataset demonstrate that our model can achieve better performance than the state-of-the-art compared models. Our source code is publicly available on GitHub \footnote{https://github.com/whonor/MINS}.

\end{abstract}

\ccsdesc[500]{Information systems~Personalization}
\ccsdesc[500]{Information systems~Recommender systems}

\keywords{News recommendation, multi-interest modeling, session-based recommendation}

\maketitle

\section{Introduction}

\label{sec:intro}
\begin{figure}[htbp]
	\centering
	\includegraphics[width=0.5\textwidth]{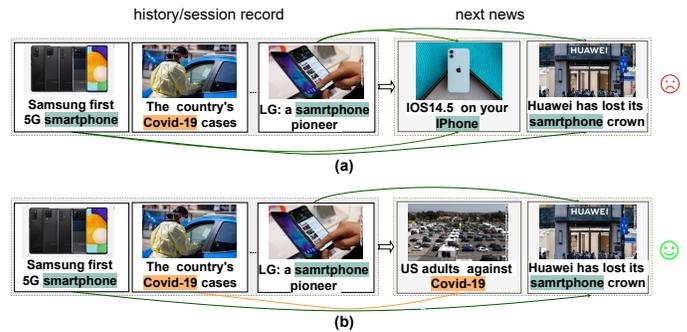}
	\caption{ Illustration of news recommender systems, among which the upper one models one interest while the lower one models two interests in the history/session. The words with the same color indicate the similar interest. 
	}
	\label{fig:example_toy}
\end{figure}
News recommendation is an important technology to help users efficiently and effectively find out their interested news from a large amount of candidate news \cite{das2007google, lian2018towards}. To model user reading interests accurately is critical for the success of news recommendation. The existing methods usually first learn user interests from their sequence of clicked/read news, then recommend the possible next news according to their relevance with user interests. For example, Wang et al. utilized knowledge graphs and knowledge-aware CNNs to model user's historical interests and candidate news to generate their representations for the subsequent recommendations  \cite{wang2018dkn}. Wu et al. proposed personalized attention networks to generate the main interest representation of the user for news recommendation \cite{wu2019npa}. Wang et al. first modeled the fine-grained user interests in text segments of different granularities to generate multi-level representations and then utilized 3D CNN to capture the relevance between user interests and candidate news \cite{wang2020fine}.

Although these existing methods have gained exceptional successes in news recommendation, they usually assume all news read/clicked in one session  share one main interest, which often violates the fact that each session may contain multiple different interests corresponding to different types of news, e.g., news from different topics. This is because a user often have diverse reading demands \cite{wu2020cprs,qi2021hie}. Taking the news session illustrated in Fig. \ref{fig:example_toy} as an example, Tom first clicked the news about \textit{Samsung's 5G smartphone} and then he was attracted to the news on \textit{Covid-19}, afterwards, he clicked another news about \textit{smartphone}. In this session, Tom's main reading interest is \textit{smartphone} revealed by the first and third piece of news and the secondary interest is \textit{Covid-19} revealed by the second piece of news. However, most of the existing methods for news recommendation work as the upper row shown in Fig. \ref{fig:example_toy}(a), which are referred as single-interest methods. They only model the user's main interest in the session while ignoring the secondary interest. As a result, they only recommend the news on \textit{smartphone} and neglect the other news that may also be of interest to the user. Different from the single-interest methods, the bottom row in Fig. \ref{fig:example_toy}(b) captures the multiple interests in the session, and recommend diverse news on both \textit{smart phone} and \textit{Covid-19} for users. Obviously, the single-interest based method fails to satisfy multiple potential interests of users, which is inferior to the multi-interest method. How to capture and represent a user's multiple interests in a news session is a critical yet challenging problem. Although multi-interest sequences have been studied in sequential/session-based recommendations~\cite{wang2019mcprn,li2019multi,cen2020controllable}, they all focus on the recommendations of product items which are quite different from news since they often do not have rich content information \cite{wang2019sequential, wang2021survey, lavie2010user, ijcaiGraphReview, wang2021hierarchical}.

Aiming at the above problem, we propose a novel \textit{multi-interest news sequence} (MINS) model for news recommendation. In MINS, a news encoder is designed to learn an informative representation of each news. 
Then a novel \textit{parallel channel interest network} (PIN) is designed to first detect the potential interest embedded in each news, and then models the multiple interests while each channel models one interest by taking those news with the same interest as the input. As a result, an informative news session representation is obtained by aggregating the interest representations from all channels. Finally, the dot production is employed to predict the next news by taking the session representation as the input.


The main contributions are summarized below:
\begin{itemize}[leftmargin=*]
	\item We propose a \textit{multi-interest news sequence} (MINS) model for news recommendation. MINS models multiple interests in a news session and recommends the next news to satisfy the diverse reading interests of each user.
	\item In MINS, a novel \textit{parallel interest network} (PIN) is devised to first detects the possible interest of each news in a session and then learns a representation for each interest from the news associated with the same interest in preparation for the subsequent next-news recommendations.    
	\item We empirically verify the effectiveness of MINS model on a real-world dataset, i.e., MIND realeased by Microsoft News \cite{wu2020mind}. Experimental results show that MINS clearly outperforms state-of-the-art news recommendation methods.
\end{itemize}

\section{Multi-interest News Sequence Learning}
\begin{figure}[t]
	\begin{center}
		\includegraphics[width=0.5\textwidth]{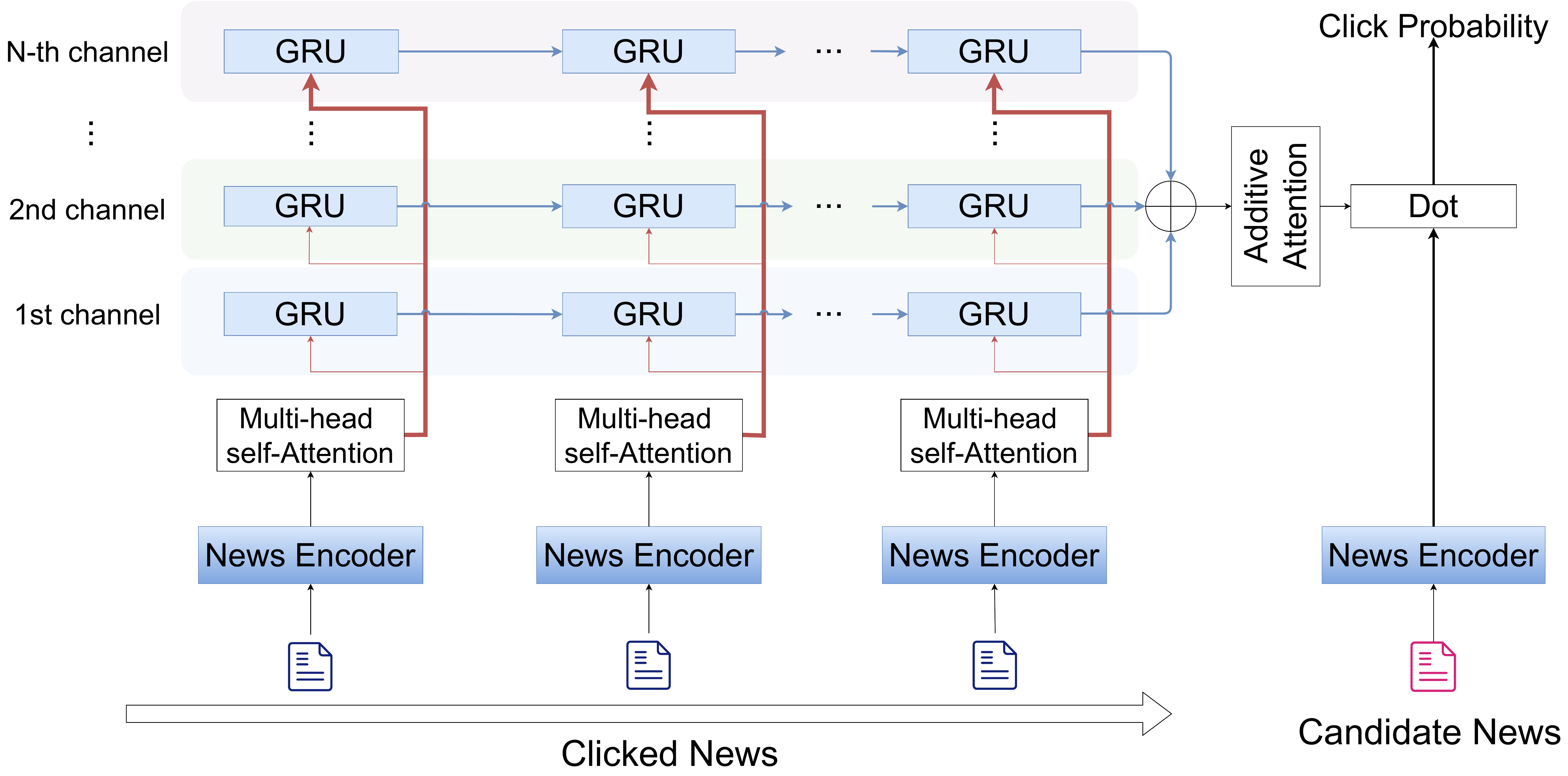}
	\end{center}
	\caption{Architecture of our MINS model.}
	\label{Fig:architeture_MISS}
\end{figure}

The architecture of our MINS model is shown in Fig. \ref{Fig:architeture_MISS}. MINS mainly contains the news encoder and the parallel-channel interest network. 

\subsection{News Encoder}
\label{sec2}

\begin{figure}[t]
	\begin{center}
		\includegraphics[width=0.35\textwidth]{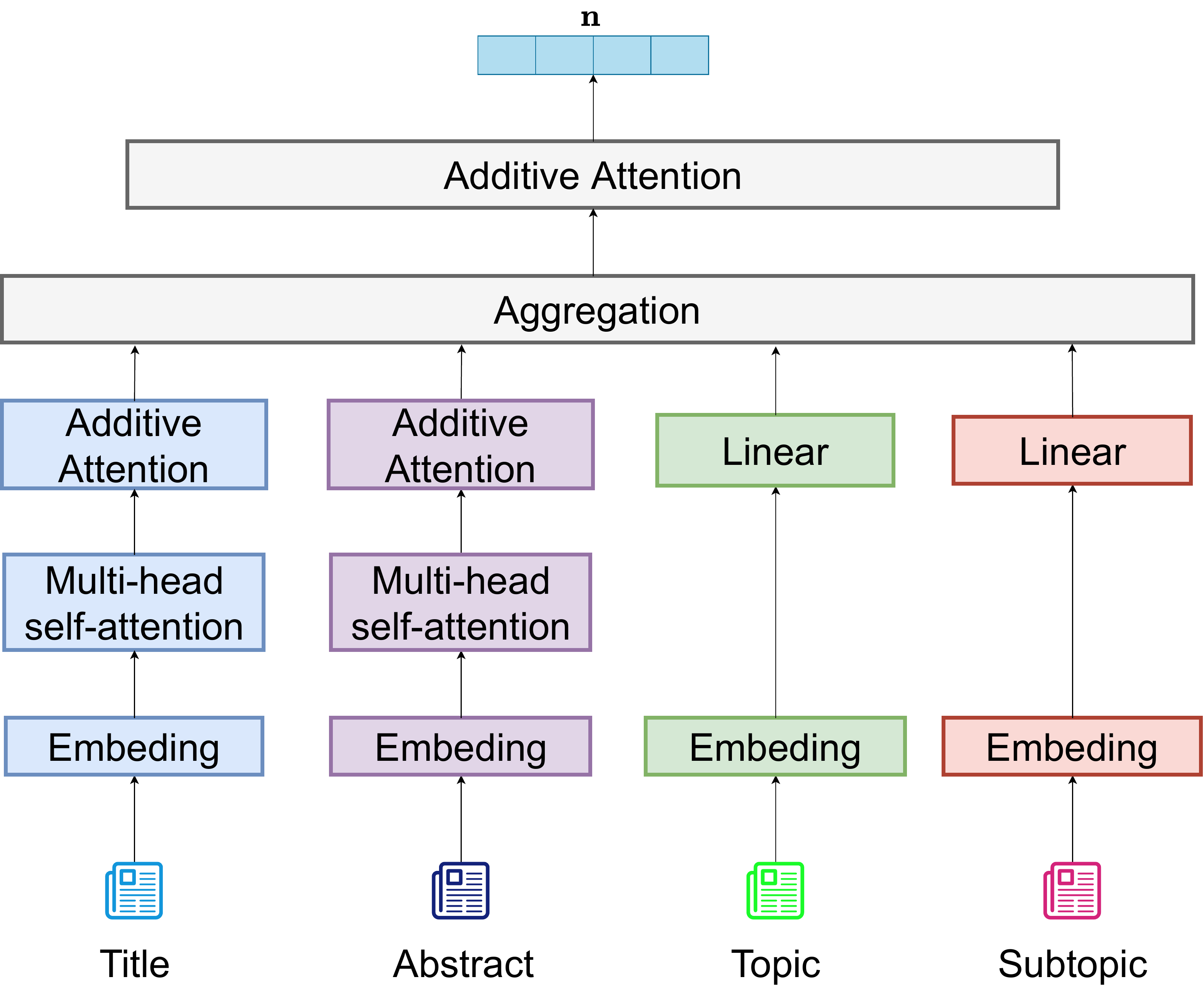}
	\end{center}
	\caption{Framework of news encoder.} \label{Fig:news encoder}
\end{figure}

To learn an informative representation for each piece of news, we design a novel news encoder. As shown in Fig. \ref{Fig:news encoder}, for each piece of news, the title, abstract, topic and subtopic are inputted into the encoder and a unified news representation $\bf{n}$ is outputted. For title and abstract, we first utilize multi-head self attention to learn the rich semantic meaning from the sentences and then employ additive attention to integrate the outputs from multi-head self attention into a unified embedding vector respectively~\cite{wu2019neurala}. We use linear transformation to encode the topic and subtopic. Finally, another additive attention module is employed to effectively aggregate the embedding vectors of all the four parts to build the final news representation $\bf{n}$.

Now we briefly introduce how to encode a news title. Given a news title with $m$ tokens denoted as $\left\{ {{w_1},{w_2}, \ldots ,{w_m}} \right\}$, it is first converted into an embedding matrix $ {\bf{E}} = \left\{ {{\bf{e}}_1,{\bf{e}}_2, \ldots ,{{\bf{e}}_m}} \right\}$ according to the pre-trained Glove embedding  \cite{pennington2014glove}, where ${\bf{e}}_m \in {{\bf{W}}_e}$, ${{\bf{W}}_e} \in {{\mathbb{R}}^{V \times D}}$, $V$ and $D$ are the vocabulary size and embedding dimension respectively. 
Inspired by Wu et al. \cite{wu2019nrms}, we further utilize multi-head self-attention \cite{vaswani2017attention} to learn a more informative word embedding based on $ {\bf{E}}$ by modelling the discriminative word-level dependencies 
, descried as below:
\begin{equation}
	\footnotesize
	\left[ {{{\bf{h}}_1},{{\bf{h}}_2},{{\bf{h}}_3}, \ldots ,{{\bf{h}}_m}} \right] = MultiHeadAttention\left( {\bf{E}} \right).
\end{equation}

To further discriminate the different importance of words and generate more informative representation for the news tile \cite{wang2018attention}, we utilize the additive attention to aggregate the aforementioned word vector representations: 
\begin{equation}
	\footnotesize
	{a_l} = {{\bf{q}}^ \top }\tanh \left( {{\bf{V}} \times {\bf{h}}_l + {\bf{b}}} \right),
	\label{Equ:add_atten_a}
\end{equation} 
\begin{equation}
	\footnotesize
	{\alpha _l} = \frac{{\exp \left( {{a_l}} \right)}}{{\sum\limits_{p = 1}^m {\exp \left( {{a_p}} \right)} }},
	\label{Equ:add_atten_alpha}
\end{equation}
where ${\bf{q}}^\top$ is the query vector, ${\bf{V}}$ and ${\bf{b}}$ are training parameters, $\alpha _l$ is the attention weight of the $l^{th}$ word in the news title.
The final title representation ${\bf{n}}^t$ is the weighted summation of word representations, i.e., ${{\bf{n}}^t} = {\sum\nolimits_{l = 1}^m {{\alpha _l}{\bf{h}}} _l}$.
Similarly, we can obtain the final abstract representation ${\bf{n}}^a$.

At the same time, we employ linear networks to generate the representations of topics and subtopics, marked as ${\bf{n}}^c$, ${\bf{n}}^{sc}$ respectively. Once the representations of the news title, abstract, topic, subtopic are ready, we concatenate them together, i.e., ${{\bf{r}}^n} = [{{\bf{n}}^t};{{\bf{n}}^a};{{\bf{n}}^c};{{\bf{n}}^s}]$. 
Then, we apply additive attentions on ${{\bf{r}}^n}$ to emphasize those important parts to generate the final news representation $\bf{n}$. The calculation is the same as shown in Eq. \eqref{Equ:add_atten_a} $\sim$ \eqref{Equ:add_atten_alpha}.  

\subsection{Parallel-channel Interest Network}
As discussed in Section \ref{sec:intro}, 
each piece of news usually embeds a particular reading interest of a user. And those news embed the same interest are often correlated while news with distinct interest are often not so related. Such observation triggers the need to model news with different interests separately to more accurately model the inter-news dependencies \cite{qi2021hie}. Accordingly, we utilize a multi-head self attention as an interest detector to detect the potential interests embedded in each news piece clicked at each time step, while each self-attention head vector represents one interest, e.g., ${\bf{d}}_i$ indicates the $i^{th}$ interest. 
The detailed operations are described below:
\begin{equation}
	\footnotesize
	[{{{\bf{d}}_1},{{\bf{d}}_2}, \ldots {{\bf{d}}_i}}\ldots, {{\bf{d}}_k}] = MultiHeadAttention\left( {\bf{n}} \right).
	\label{Equ:multi-head}
\end{equation}

Then, we devise a multi-channel GRU-based recurrent network where each channel models the sequential dependencies over news within each interest. Finally we take the hidden state at the final step as the representation of each interest. Here the number of channels $k$ is empirically set to 6 for best performance, which is consistent with the number of interests, i.e., the number of heads in Eq (\ref{Equ:multi-head}). For the $i^{th}$ channel, in each step, the GRU cell is updated as follows:     
\begin{equation}
	\footnotesize
	{{\bf{g}}_t} = {\sigma _s}\left( {{{\bf{W}}_g}\left[ {{{\bf{d}}_{i,t}},{{\bf{x}}_{t - 1}}} \right]} \right),
\end{equation}
\begin{equation}
	\footnotesize
	{{\bf{z}}_t} = {\sigma _s}\left( {{{\bf{W}}_z}\left[ {{{\bf{d}}_{i,t}},{{\bf{x}}_{t - 1}}} \right]} \right),
\end{equation}
\begin{equation}
	\footnotesize
	{{\bf{u}}^\prime } = {\sigma _t}\left( {{{\bf{W}}_h}\left[ {{{\bf{d}}_{i,t}},{{\bf{x}}_{t - 1}} \odot {{\bf{g}}_t} } \right]} \right),
\end{equation}
\begin{equation}
	\footnotesize
	{{\bf{u}}_t} = \left( {1 - {{\bf{z}}_t}} \right) \odot {{\bf{x}}_{t - 1}} + {{\bf{z}}_t} \odot {{\bf{u}}^\prime },
\end{equation}
where ${\bf{g}}_t$ and ${\bf{z}}_t$ are the reset gate and update gate in GRU \cite{Hidasi2016gru4rec}. ${\bf{d}}_{i,t}$ is the $i^{th}$ interest representation extracted from the news piece clicked at the time $t$ by using Eq (\ref{Equ:multi-head}). ${\bf{x}}_{t-1}$ is the last interest state in the $i^{th}$ channel, $\sigma _s$ and $\sigma _t$ are activation functions, which are specified as sigmoid and tanh respectively. 

Following this process in one channel till to the last step $t$, we can obtain the user's current $i^{th}$ interest representation ${\bf{u}}_t$ by taking the interest representation ${\bf{d}}_i$ from each step as the input. 
Further, we integrate ${\bf{u}}_t$ from all the $k$ channels together to obtain the user's final compound interest representation. Specifically, additive attention is employed to emphasize the crucial information from the multiple interest representations and its calculation is presented in Eq. \eqref{Equ:add_atten_a} $\sim$ \eqref{Equ:add_atten_alpha}. 

Therefore, the final news session representation embedding the user's diverse reading interests is calculated as a attention-based weighted sum of the outputs from all the $k$ channels: 
\begin{equation}
	\footnotesize
	{\bf{s}} = \sum\nolimits_{i = 1}^k {{\beta _i}{\bf{u}}_{t,i}},
\end{equation}
where $\beta_i$ is the aggregation weight for the $i^{th}$ channel calculated in the same way as described in Eq.\eqref{Equ:add_atten_a} $\sim$ \eqref{Equ:add_atten_alpha}.

\subsection{Click Predictor and Optimization}

Given a session consisting of a sequence of clicked news, our MINS model employs the parallel-channel interest network to effectively capture the multiple interests contained in the news session to generate the session representation, marked as $\bf{s}$. For a piece of candidate news, MINS utilizes the news encoder to generate the news representation, marked as $\bf{c}$. The probability $\hat y$ of the user will click/read the candidate news is computed as the dot production of $\bf{s}$ and $\bf{c}$, i.e., $\hat y = {\bf{s}^\top}\cdot{\bf{c}} $ \cite{okura2017embedding}. 

Motivated by An et al. \cite{An2019LSTUR}, the task of news recommendation can be regarded as a pseudo $K$ + 1-way classification problem.
Hence, we apply a mini-batch gradient descent on the log-likelihood loss to train our model \cite{huang2013learning}. The loss function is as follow: 
\begin{equation}
	\footnotesize
	{L} =  - \sum\limits_{j = 1}^P {\log \frac{{\exp \left( {\hat y_j^ + } \right)}}{{\exp \left( {\hat y_j^ + } \right) + \sum\nolimits_{n = 1}^K {\exp \left( {\hat y_{j,n}^ - } \right)} }}} ,
	\label{e22:lossnr}
\end{equation}
where \(P\) is the number of positive samples, $K$ is the number of negative samples. ${\hat y_j^ + }$ is the click score of the $j^{th}$ positive sample, ${\hat y_{j,n}^ - }$ is the click score of the $n^{th}$ negative samples w.r.t. the $j^{th}$ positive sample in the same session.

\begin{table*}[htbp]
	\footnotesize
	\centering
	\caption{Comparison result with baseline methods.}
	\begin{tabular}{c|cccccccc}
		\toprule
		\multirow{2}[4]{*}{Model} & \multicolumn{4}{c|}{MIND-small} & \multicolumn{4}{c}{MIND-large} \\
		\cmidrule{2-9}          & AUC   & MRR   & nDCG@5 & \multicolumn{1}{c|}{nDCG@10} & AUC   & MRR   & nDCG@5 & nDCG@10 \\
		\midrule
		BiasMF & 0.5108  & 0.2258  & 0.2318  & \multicolumn{1}{c|}{0.2952 } & 0.5111  & 0.2257  & 0.2346  & 0.2963  \\
		DKN   & 0.5726  & 0.2339  & 0.2418  & \multicolumn{1}{c|}{0.3033 } & 0.6329  & 0.2902  & 0.3163  & 0.3930  \\
		LSTUR & 0.6021  & 0.2659  & 0.2873  & \multicolumn{1}{c|}{0.3529 } & 0.5633  & 0.2454  & 0.2583  & 0.3252  \\
		NRMS  & 0.6391  & 0.3017  & 0.3282  & \multicolumn{1}{c|}{0.3937 } & \underline{0.6701}  & \underline{0.3185}  & \underline{0.3534}  & \underline{0.4175}  \\
		HiFi-Ark & 0.6403  & 0.2996  & 0.3272  & \multicolumn{1}{c|}{0.3925 } & 0.6394  & 0.2969  & 0.3221  & 0.3888  \\
		TANR  & \underline{0.6455}  & \underline{0.3107}  & \underline{0.3367}  & \multicolumn{1}{c|}{\underline{0.4017} } & 0.6611  & 0.3148  & 0.3467  & 0.4114  \\
		MINS  & \textbf{0.6710 } & \textbf{0.3171 } & \textbf{0.3525 } & \multicolumn{1}{c|}{\textbf{0.4150 }} & \textbf{0.6811 } & \textbf{0.3249 } & \textbf{0.3601 } & \textbf{0.4242 } \\
		\midrule
		Improvement$^1$ (\%) & 3.95  & 2.06  & 4.69  & 3.31  & 1.64  & 2.01  & 1.90  & 1.60  \\
		\bottomrule
	\end{tabular}%
	\label{tab2}%
	\begin{tablenotes}
		\footnotesize
		\item[1] $^{1}$ \small{Improvement achieved by MINS over the best-performing baselines (TANR and NRMS on MIND-small and MIND-large respectively).}
	\end{tablenotes}
\end{table*}%

\section{Experiments and Evaluation}
\subsection{Experimental Setup and Baselines}
Our MINS model can achieve the best performance with following parameters. The dimension of word embedding is set to 300. The learning rate is 0.0001. The number of self-attention heads in multi-head self-attention of news encoders is set to 15. In parallel-interest networks, the number of heads in multi-head self-attention is set to 6, and the number of GRU layers, i.e, interest channels, is also set to 6. We employ the pre-trained Glove embedding \cite{pennington2014glove} to initialize word embedding matrix.
Since there is no test set released for the MIND dataset from Microsoft News, we split \({\rm{10\% }}\) samples from the train set as the validation set, and take the released validation set as the test set.

We compare our model with following representative and state-of-the-art baselines, including BiasMF \cite{koren2009matrix}, DKN \cite{wang2018dkn}, Hi-Fi Ark \cite{Liu2019Hi}, TANR \cite{wu2019neurala}, NRMS \cite{wu2019nrms} and LSTUR \cite{An2019LSTUR}. Following the work of Wu et al. \cite{wu2020cprs}, we apply four metrics i.e. AUC, MRR, nDCG@5 and nDCG@10 to evaluate their performance.

\subsection{Comparison with Baseline Methods}
The experimental comparison results on MIND dataset are shown in Table \ref{tab2}. There are several observations. First, compared with neural network methods, the method based on statistical machine learning such as BiasMF consistently shows the worse performance on two versions of MIND dataset. This is because that BiasMF fails to capture nonlinear features and complex semantic representations for news recommendation. 
Second, the method applying multi-head self-attention such as NRMS outperforms the methods without multi-head self-attention(i.e., DKN, LSTUR, HiFi-Ark, TANR) on the MIND-large dataset.
This may because multi-head self-attention can efficiently model news content representations.
Third, our model significantly outperforms other baselines on all metrics and datasets, which improves about 4$\%$ in terms of AUC on MIND-small dataset than TANR. The reason may be that MINS tends to learn multiple interests contained in news session, which can satisfy the diverse requirements of users.

\subsection{Ablation Study}

\begin{figure}[htbp]
	\centering
	\begin{minipage}{0.49\linewidth}
		\centering
		\includegraphics[width=\linewidth]{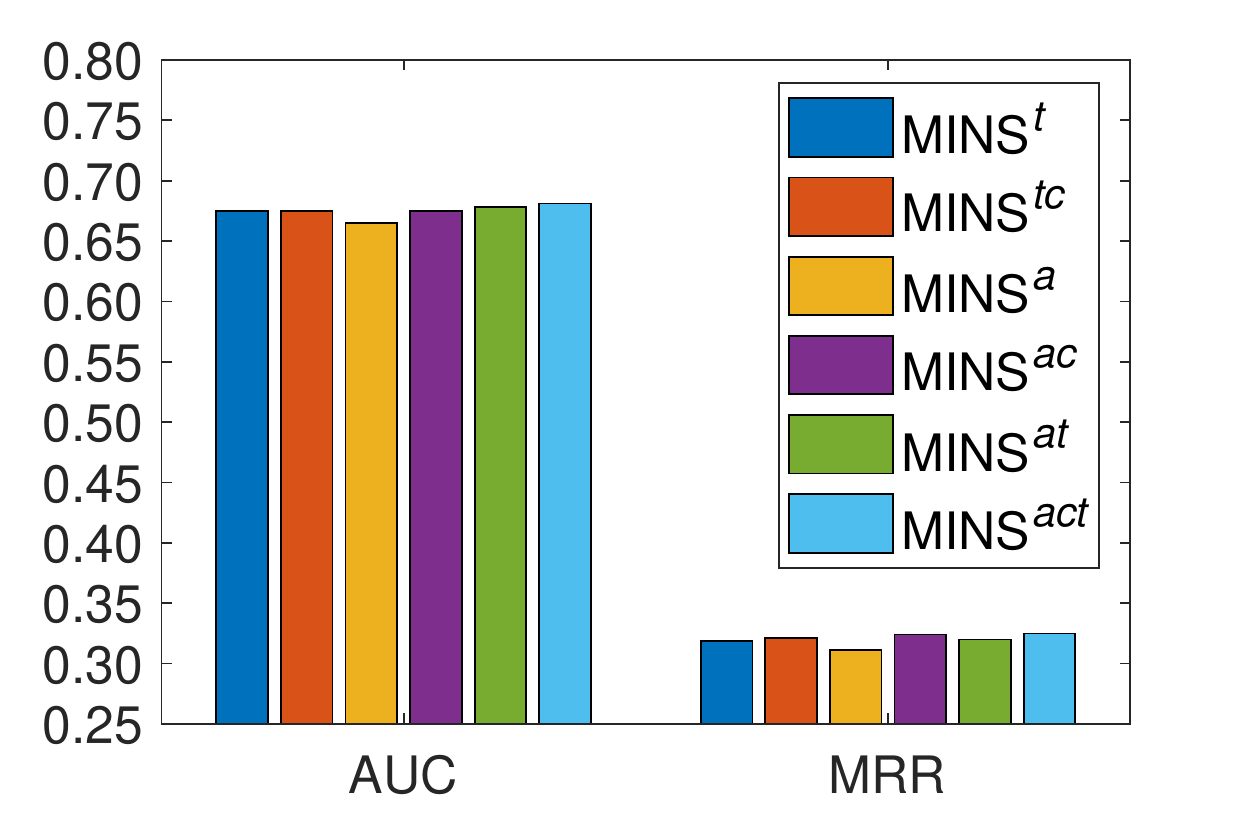}
		\caption{Impact of inputs.}
		\label{input}
	\end{minipage}
	\begin{minipage}{0.5\linewidth}
		\centering
		\includegraphics[width=\linewidth]{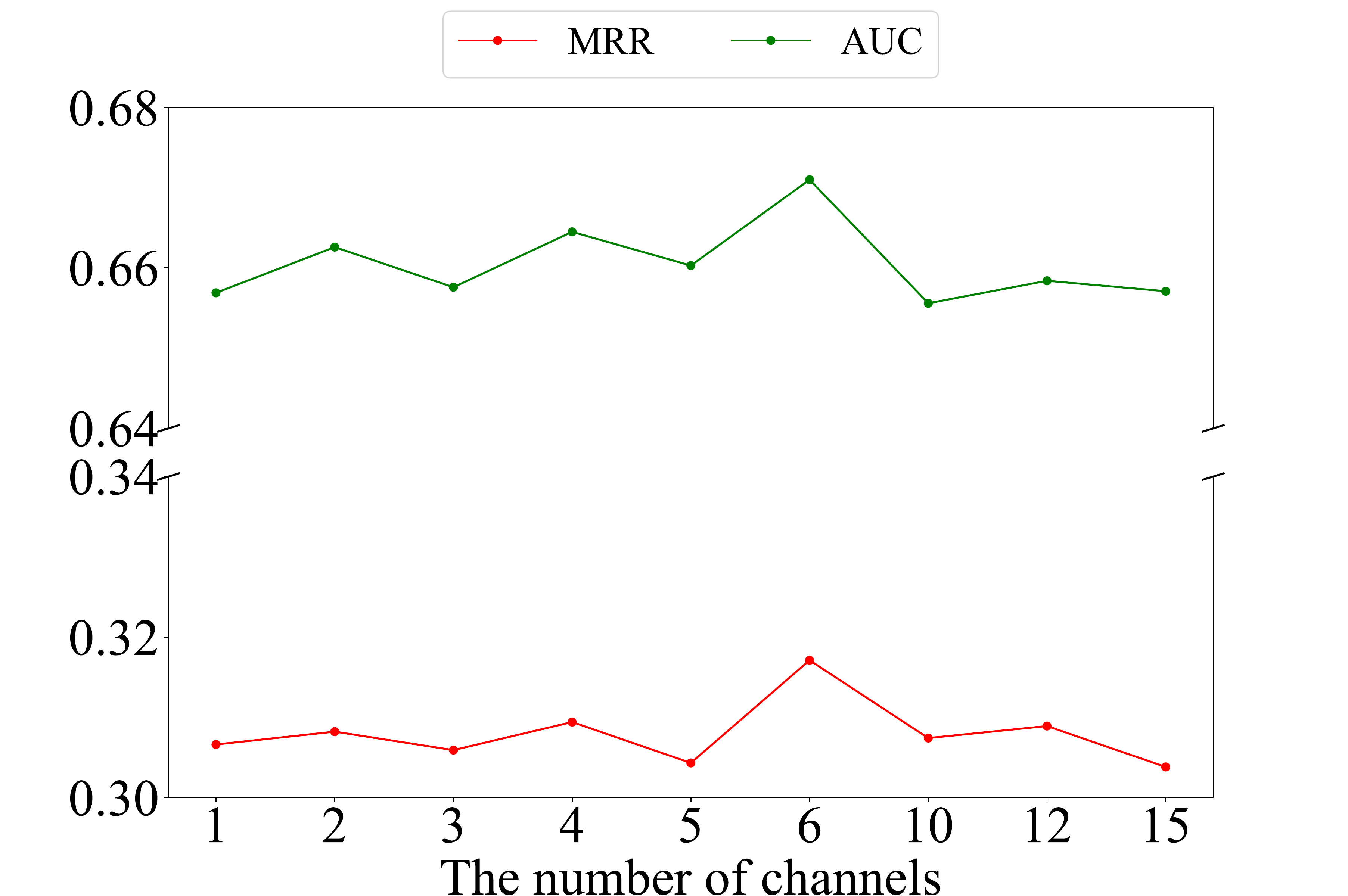}
		\caption{Impact of channels.}
		\label{layer}
	\end{minipage}
\end{figure}

\textbf{Impact of input data.} In order to evaluate the effectiveness of different input data, we combine titles, abstracts, topics and subtopics and feed them into our model. The standard MINS are transformed into six variants, that is, MINS$^{t}$, MINS$^{tc}$, MINS$^{a}$, MINS$^{ac}$, MINS$^{at}$ and MINS$^{act}$ based on the difference of input data. We respectively denote the superscript $t$, $a$ and $c$ as titles, abstracts, the group of topic and subtopic. The comparison results are shown in Fig.  \ref{input}. According to the figure, we found MINS$^{act}$ achieve the best performance, which demonstrates that it is reasonable and necessary for our model to integrate the four kinds of news information.

\textbf{Impact of GRU/interest channel.} 
For our MINS model, GRU channels in the parallel-interest network is used to model the multiple interests in the session, which is essential for MINS to achieve better performance. Due to the limitation of model structure, the possible number of channels must be integer factors of the word embedding dimension (300). We investigate the effectiveness with different numbers of channels, as shown in Fig. \ref{layer}. According to the figure, we find that when the number of channels is set to 6, our model can achieve the best performance. The less channels may fail to capture the diverse interests while the more channel may lead to overfitting issue.

\section{Conclusion}   
In this paper, we propose a \textit{multi-interest news sequence} (MINS) model for news recommendation. A parallel-interest network is devised to detect the potential interest of each news and assign it into the corresponding interest-specific channel, followed by GRU based network to generate the multi-interest representation for the session. Besides, a news encoder is devised to learn the accurate news representation with multi-head self-attentions. 
Our extensive experiments demonstrate that our MINS model can outperform the state-of-the-art compared baselines for news recommendation. In future, we will attempt to utilize more powerful pre-trained language models and knowledge graphs to further improve the performance of our MINS model.

\begin{acks}
The research work is partly supported by National Natural Science Foundation of China under Grant No.61502259 and No.11901325, National Key R\&D Program of China under Grant No.2018YFC0831700, and Key Program of Science and Technology of Shandong Province under Grant No.2020CXGC010901 and No.2019JZZY020124.
\end{acks}

\bibliographystyle{unsrt}
\bibliographystyle{ACM-Reference-Format}
\bibliography{ref}

\appendix

\end{document}